# Linear Trimer Formation by Three-Center–Four-Electron Bonding in RuP

Daigorou Hirai*,†, Keita Kojima‡, Naoyuki Katayama‡, Mitsuaki Kawamura†,

Daisuke Nishio-Hamane†, and Zenji Hiroi*,†

†*Institute for Solid State Physics, University of Tokyo, Kashiwa, Chiba 277-8581, Japan*

‡*Department of Applied Physics, Nagoya University, Nagoya 464–8603, Japan*

**ABSTRACT:** In molecules like hydrogen, most chemical bonds are formed by sharing two electrons from each atom in the bonding molecular orbital (two-center–two-electron (2c2e) bonding). There are, however, different kinds of chemical bonding. The $I_3^-$ molecule, for example, is noteworthy because three iodine atoms are linearly united by sharing four electrons (three-center–four-electron (3c4e) bonding). Some inorganic solids undergo phase transitions that result in the formation of "molecules" in their crystalline frameworks, which are often accompanied by dramatic changes in physical properties; the metal-to-insulator transition (MIT) in vanadium dioxide, for example, occurs with the formation of dimer molecules with 2c2e bonding. We repot the creation of a linear ruthenium trimer with 3c4e bonding in ruthenium monopnictide at its MIT. Charge transfer from polymerized phosphorous to ruthenium produces this unusual molecule, with all conduction electrons trapped by the bonding molecular orbital. Our results demonstrate that molecules are crucial even in solid crystals as they impact their electronic properties.

## INTRODUCTION

The most typical chemical bonds are formed by sharing two electrons from each atom in the bonding molecular orbital [two-center–two-electron (2c2e) bonding]. On the other hand, chemical bonds in molecules are diverse, and some chemical bonds are formed by the sharing of electrons between two or more nuclei. The simplest example is $H_3^+$, where three H nuclei share two electrons to form a three-center–two-electron (3c2e) bond to form an equilateral triangular $H_3^+$ molecule[1–3]. Another notable example is $I_3^-$ or $HF_2^-$, where a linear trimer of nuclei shares four electrons (3c4e bonding) to form a molecule[4,5]. Chemical bonds are formed even in solids, which are usually followed by significant changes in physical properties; for example, when dimer molecules with 2c2e bonding formed in vanadium dioxide metal-to-insulator transition (MIT) occurs[6,7].

In this study, we report the observation of the formation of three-center–four-electron (3c4e) bonding in RuP at the MIT. Transition metal pnictides have been studied for a long time because of their intriguing electronic phases[8]. For example, spin/orbital density waves were observed in MnP[9], showing a topological Hall effect arising from helimagnetic ordering[10]. Furthermore, suppression of helimagnetic ordering by pressure in CrAs and MnP leads to superconductivity around the quantum critical point (QCP)[11–13], while superconductivity in WP was observed at ambient pressure[14]. MnP attracted attention as a topological material[15], and the Wyle semimetals TaAs, NbP, and FeP exhibited unusual linear magnetoresistance[16–18].

Ruthenium pnictides RuPn (Pn = P, As) also have remarkable characteristics[19,20]. They show two phase transitions: from paramagnetic metal to a pseudogap (PG) phase at $T_{PG}$ = 330 K (250 K) and to a nonmagnetic insulator at $T_{MI}$ = 270 K (190 K) for RuP (RuAs)[21]. The successive phase transitions are gradually suppressed by substituting Rh for Ru, and a dome-shaped superconducting phase emerges around the QCP, where the PG phase is suppressed to 0 K. This superconductivity is suggested to be a full-gap BCS type.[22-24]. The superconducting pairing mechanism should be related with certain fluctuations in the parent phase, however the superconducting mechanism remains unclear due to the unknown mechanism of the MIT.

The MIT in RuPn has been attributed to a charge density wave (CDW) instability [25–27] or molecular orbital formation of Ru[21,28], although neither has been evidently supported. At low temperatures, RuAs have a 3×3×3 superlattice structure, while the low-temperature structure of RuP is unidentified. In both RuAs and RuP, there is no nesting of Fermi surfaces corresponding to this or other superlattice structure[26], casting doubt on the CDW mechanism. Furthermore, it is difficult to explain how the CDW fully gaps the three-dimensional Fermi surfaces. First-principles calculations, on the other hand, point to the molecular formation of $4d_{xy}$ orbitals of ruthenium, which have a dominant contribution around the Fermi energy ($E_F$)[28,29]. Although the nonmagnetic insulator is explained by the formation of molecular orbital, no molecular orbital arrangement pattern consistent with the low-temperature superlattice structure has been revealed in both compounds. Furthermore, RuP shows a cumbersome sample dependence, with polycrystalline samples becoming nonmagnetic insulators at low temperatures[21], while single crystals re-maining metallic[25]. In order to elucidate the mechanism of the MIT in RuPn, the intrinsic properties of RuP and its crystal structure at low temperatures must be revealed.

We examined polycrystalline and single crystal samples of RuP. Compositional analysis revealed that the polycrystalline sample was nearly stoichiometric, whereas the single crystal sample had a phosphorus deficiency of approximately 10%, which doped electrons and produced finite Fermi surfaces even below $T_{MI}$. Synchrotron X-ray diffraction (XRD) experiments on polycrystalline samples revealed the formation of a 3×3×3 superstructure in the nonmagnetic insulator phase, which is analogous to RuAs. The formation of a linear trimer of Ru was found to be the origin of the superstructure, and the development of the $d_{xy}$ orbital trimer drives the MIT, according to the first-principles calculations. We propose that a three-center–four-electron (3c4e) bond is created by 4/3 electrons per $d_{xy}$ orbital, which is more than the one expected from the formal valence $Ru^{3+}$ due to electron transfer from phosphorous to ruthenium.

**EXPERIMENTAL SECTION**

Polycrystalline and single crystal samples were synthesized by the conventional solid-state reaction and the tin flux method, respectively, as previously reported[21,25]. XRD measurements of a RuP polycrystalline sample at 400, 280, and 100 K were carried out using synchrotron X-ray beams with a wavelength of $\lambda$ = 0.414 Å at 11-BM in Advanced Photon Source (Argonne National Laboratory). Rietveld analysis was performed by Rietan-FP[30]. Compositional analysis was performed by energy dispersive X-ray spectroscopy in a scanning electron microscope (JEOL JSM-5600). Ru:P ratios were 1.000(2):1.023(2) and 1.000(5):0.910(3) for polycrystalline and single crystal samples, respectively. The polycrystalline sample is nearly stoichiometric, whereas the single crystal sample is approximately 10% deficient in phosphorous relative to ruthenium. Single crystals with low phosphorous concentration were always obtained within our experimental conditions. Magnetic susceptibility was measured in a magnetic properties measurement system (MPMS-3, Quantum Design). Electrical resistivity, Hall effect, and heat capacity measurements were performed in a physical properties measurement system (PPMS, Quantum Design).

**RESULTS AND DISCUSSION**

Figure 1(a) compares the temperature dependence of the resistivity of a RuP polycrystalline sample and a RuP$_{1-x}$ single crystal. Both show an increase below $T_{MI}$ = 270 K from the metallic state at high temperature. The RuP polycrystalline sample behaves as an insulator at low temperatures, whereas the RuP$_{1-x}$ single crystal shows a smaller increase at $T_{MI}$, followed by metallic conduction below ~200 K. The latter is qualitatively and quantitatively consistent with "RuP" single crystals reported by other groups[25]. The differences between the two also appear in the heat capacity and Hall coefficient (Supplementary information Figs. S1, S2). Sommerfeld coefficients ($\gamma$) for polycrystalline and single crystal samples are 0.32(1) mJ K$^{-2}$ mol$^{-1}$ and 1.23(1) mJ K$^{-2}$ mol$^{-1}$, respectively, indicating that the single crystal has a finite density of states (DOS) at $E_F$. The Hall coefficients at the lowest temperature are high positive and small negative, respectively, demonstrating a metallic state with electron carriers in the single crystal, which is probably caused by a phosphorous deficit.

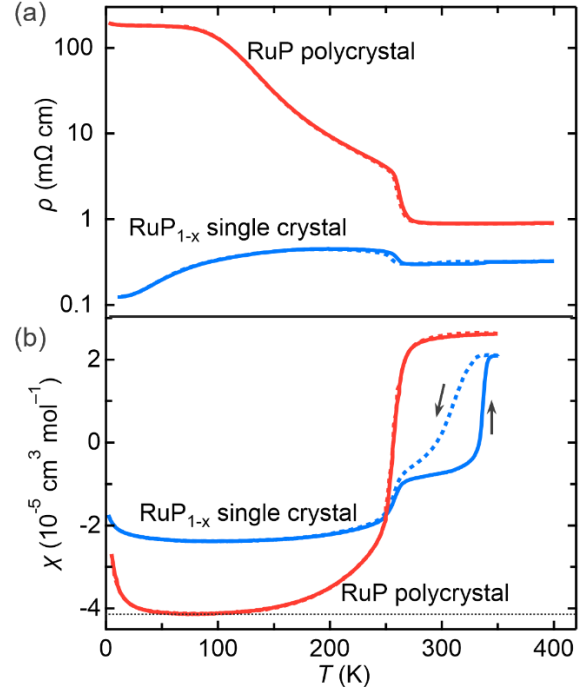

**Figure 1.** Temperature dependences of (a) electrical resistivity and (b) magnetic susceptibility ($B$ = 7 T) for a polycrystalline sample of RuP (red lines) and a single crystal of RuP$_{1-x}$ (blue lines). The solid and broken lines indicate the data measured on heating and cooling, respectively.

The magnetic susceptibility $\chi(T)$ of the polycrystalline sample is temperature independent above $T_{MI}$ = 270 K, confirming Pauli paramagnetic metal [Fig. 1(b)]. Below $T_{MI}$, $\chi(T)$ steeply drops to $-4.1\times10^{-5}$ cm$^3$ mol$^{-1}$ and is temperature independent, except for a Curie tail below 40 K. The spin susceptibility is $-0.2\times10^{-5}$ cm$^3$ mol$^{-1}$, nearly zero, after subtracting the orbital paramagnetic component of $4.9\times10^{-5}$ cm$^3$ mol$^{-1}$[31] and the diamagnetic contribution of the nuclear diamagnetism of $-8.8\times10^{-5}$ cm$^3$ mol$^{-1}$, suggesting a nonmagnetic state. The $\chi(T)$ of the single crystal sample, on the other hand, drops in two steps at $T_{PG}$ = 330 K and $T_{MI}$ = 270 K, from a small temperature-independent value at high temperatures. At low temperatures, the $\chi(T)$ is higher than for the polycrystalline sample, implying that single crystal has a Pauli paramagnetic component of $\chi_P$ = $1.5\times10^{-5}$ cm$^3$ mol$^{-1}$. This $\chi_P$ value agrees well with the $1.7\times10^{-5}$ cm$^3$ mol$^{-1}$ calculated from $\gamma$ by assuming free electrons. With the conduction electrons induced by phosphorous deficiency, RuP$_{1-x}$ single crystal stays a Pauli paramagnetic metal below 270 K. The previously reported sample-dependence is caused by phosphorous deficiency, and the intrinsic properties of RuP manifest in polycrystalline samples.

We performed synchrotron XRD experiments on a polycrystalline sample of RuP to reveal the structural change caused by the MIT. Figure 2(a) shows XRD patterns corresponding to the metallic, PG, and insulator phase at 400, 280, and 100 K. The 400 K pattern is consistent with a MnP-type structure, as previously described[21] (SI, Fig. S3, Table S1). The 280 K and 400 K patterns are almost identical, however at 100 K, several superlattice reflections with a maximum intensity of approximately 1% of the strongest reflection appear [inset of Fig. 2(a)]. The lowest-angle superlattice reflection can be indexed as 1/3 1/3 1/3 based on the high-temperature orthorhombic unit cell, and the other superlattice reflections can be indexed as their higher-order reflections. Thus, at low temperatures, a superlattice structure 3×3×3 times larger than the high-temperature unit cell is formed. Because this superlattice is identical to that of the insulator phase of RuAs[26], RuP probably experiences similar lattice deformation.

Rietveld analysis was performed for the XRD pattern at 100 K assuming a 3×3×3 superlattice [Fig. 2(b)]. The initial structure for the refinement was the low-temperature monoclinic structure of RuAs[26]. There were 9 general sites each for Ru and P, with a total of 27 coordinates refined. The Debye–Waller factors for all Ru sites or all P sites were retained the same, and the occupancy of all elemental sites was set to 1. The goodness-of-fit parameter $S$ is 1.0484, demonstrating the satisfactory refinement (SI, Table S2). In particular, the intensity of the superlattice reflections is accurately reproduced, as shown in the inset of Fig. 2(b).

The Ru–Ru distance undergoes the most significant structural change through the MIT. At 400 K, a zigzag chain of Ru extends in the $b$ direction, with Ru–Ru lengths of 3.172 and 2.970 Å in the leg and lung directions, respectively [Fig. 3(a)]. At 100 K, the Ru–Ru distance in the leg direction is significantly modulated, resulting in the formation of a linear trimer. (SI, Fig. S5). Compared to 400 K, the bond inside the trimer shrinks by approximately 5%, whereas the distance between trimers expands by more than 10% (SI, Table S3). The bond modulation in the $b$ direction is considerable (17%), while the bond distance between chains changes only little (3.7%). The linear trimers are aligned with phase shift in both the $c$ and $a$ axes, resulting in a 3×3×3 superlattice. Therefore, the formation of the Ru linear trimer drives the structural phase transition. Ru has a displacement of 0.18 Å, which is substantially larger than the atomic displacements in the CDW transition of other materials[32]. For example, in 2H-TaSe$_2$, Ta is displaced by 0.085 Å[33], and in 1T-TiSe$_2$, Ti is displaced by 0.05 Å[34] upon the CDW transition. Thus, MIT is most likely caused by a mechanism other than simple CDW.

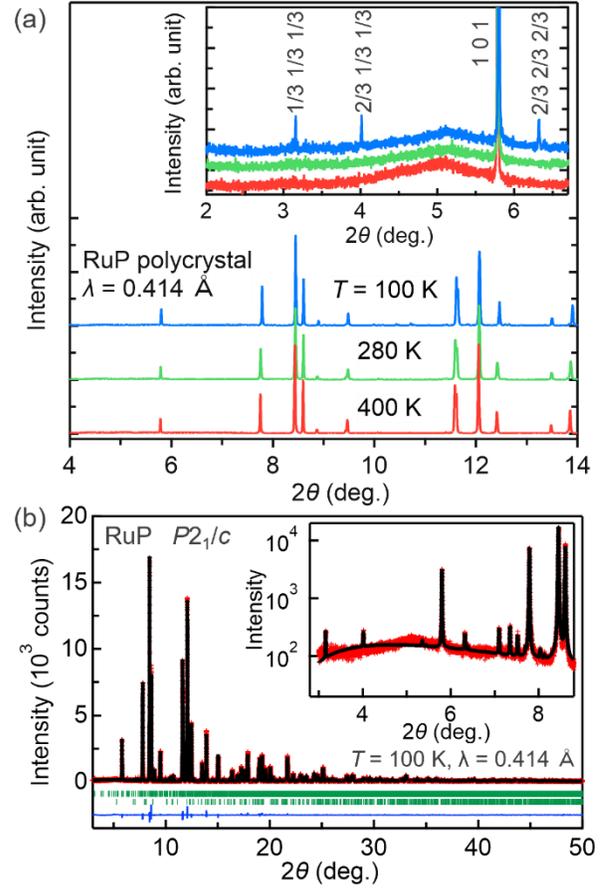

**Figure 2.** (a) XRD patterns of a polycrystalline sample of RuP at 400, 280, and 100 K corresponding to the metallic, pseudo-gap, and insulating phases, respectively. The inset in (a) shows the appearance of superlattice reflections at 100 K. (b) Rietveld fitting for the 100 K data. Observed (red circles), calculated (black solid line), and difference (lower blue solid line) XRD patterns are shown. Green tick marks indicate the position of allowed reflections. The inset in (b) is the logarithmic scale plot at the low angle range.

To elucidate the change in the electronic structure at the MIT, first-principles calculations were performed using the obtained structural parameters without considering the electronic correlation (details in SI). Figure 4 compares the DOS for the high- and low-temperature phases. The DOS for the high-temperature phase shows a sharp peak around the $E_F$ as reported previously[28]. In the low-temperature phase, the peak disappears, creating an energy gap of approximately 0.06 eV. The opening of the band gap without considering electron correlation indicates that the insulator phase is a band insulator rather than a Mott insulator.

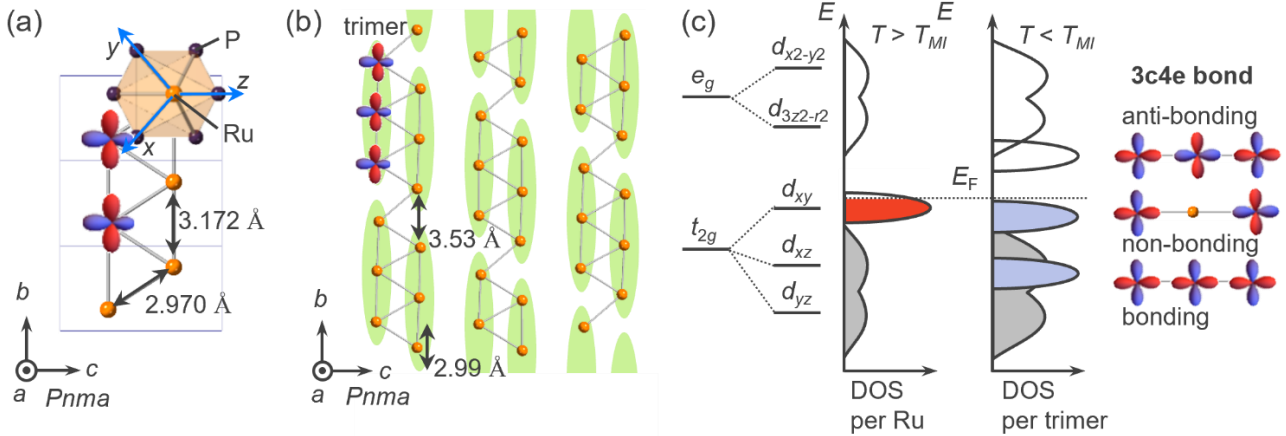

**Figure 3.** (a) Zigzag chain of Ru atoms in the high-temperature structure. The $d_{xy}$ orbitals (red and blue clovers) are arranged to form a chain along the $b$ axis with significant overlap with their neighbors. The chain is uniform with a Ru–Ru bond length of 3.172 Å. (b) Linear trimer (green ellipse) formation along the chain in the insulator phase. The Ru–Ru distance is modulated to have shorter bonds in the trimers (2.99 Å) and longer bonds between the trimers (3.53 Å) in a chain. (c) Schematic energy diagram of the $d$ states in RuP around the $E_F$. The $d_{xy}$ state forms a narrow quasi-one-dimensional band, while the other states form broad bands. The $E_F$ is located at the partially filled $d_{xy}$ band at high temperatures above $T_{MI}$. When the linear orbital trimer is produced at low temperatures, the $d_{xy}$ band splits into bonding, non-bonding, and antibonding bands that are derived from three different types of $d_{xy}$ orbital arrangements, as depicted on the right. Between the occupied non-bonding band and the vacant antibonding band, an energy gap emerges. This indicates that the linear trimer is stabilized by three-center–four-electron (3c4e) bonding.

The $t_{2g}$ orbitals of Ru are primarily responsible for the states in the vicinity of $E_F$. In the high-temperature phase, the distorted RuP$_6$ octahedral crystal field splits the $t_{2g}$ levels into $d_{yz}$, $d_{xz}$, and $d_{xy}$ orbitals in order from low energy, as illustrated in Fig. 3(a)[28]. Among them, the $d_{xy}$ orbital state dominates the DOS near $E_F$. according to the photoemission spectroscopy experiments[29]. In the $b$ direction, where the lobe extends, $d_{xy}$ orbitals have a substantial overlap with neighboring orbitals, whereas the transfer in the lung direction must be small. As a result, the $d_{xy}$ orbitals form a chain along the $b$ axis, and their one-dimensional band has a sharp DOS peak around $E_F$, inducing electronic instability[28].

The $d_{xy}$ band is filled with more electrons than expected from the formal charges, according to calculations for the high-temperature phase. This is probably due to charge transfer from phosphorous to ruthenium, as evidenced by photoemission spectroscopy[29]. At high temperatures, RuP has a MnP-type structure, which is a derivative of the NiAs-type structure[35,36]; VP with the NiAs-type structure has an equivalent spacing between phosphorous atoms of 3.180(5) Å[37]. In RuP at 400 K, on the other hand, the zigzag chain of phosphorus with a short bond of 2.757(1) Å extends in the $b$ direction with a large distance of 3.855(1) Å between the chains, indicating a strong covalent bond between the 3p orbitals of phosphorous in the zigzag chain. In such polyanionic compounds, the octet rule is generally violated, resulting in a decrease in the formal charge of the anions[38]. In other words, the covalent bond polymerizes the P$^{3-}$ ion to produce the P$^{(3-\delta)-}$ ion, which subsequently transfers the excess electrons to ruthenium.

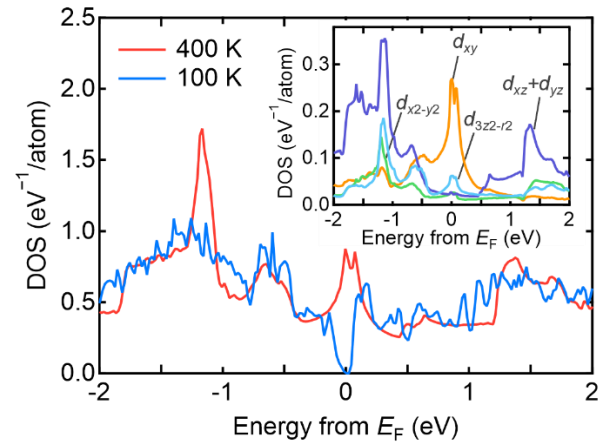

**Figure 4.** (a) DOS of RuP obtained by first-principles calculations for the metallic phase at 400 K (red line) and insulator phase at 100 K (blue line). The Fermi energy of the metallic phase is set to zero. The inset shows the partial DOS of five Ru 4$d$ states for the metallic phase.

When a linear trimer of ruthenium is formed [Fig. 3(b)], the three $d_{xy}$ orbitals create an orbital cluster, in this case an orbital trimer, by σ-bonding, as shown schematically in Fig. 3(c). As a result, three molecular orbitals are formed depending on the phase of the atomic wave function of the three $d_{xy}$ orbitals: bonding, nonbonding, and antibonding. The bonding and nonbonding orbitals are filled, resulting in a band insulator with an energy gap below an empty antibonding state. Note that, because each trimer has four electrons, the linear trimer is stabilized by 3c4e bond. The number of occupied $4d$ electrons in ruthenium could not be evaluated from the band calculation due to the significant hybridization between ruthenium and phosphorous. In RuAs with smaller hybridization, on the other hand, the formal valence of ruthenium in the insulator phase was evaluated to be approximately 2.7[26], which is consistent with the aforementioned 3c4e bonding. The number of electrons transferred from phosphorous to ruthenium in the high-temperature phase is unlikely to be equal to 1/3 by coincidence, and it is considered to be locked to 1/3 at the MIT to match the lattice periodicity (commensurability). In fact, incommensurate superlattice reflections have been observed in RuAs in the intermediate PG phase[26], indicating that an array of orbital clusters with a period incommensurate with the lattice is likely to form. The subsequent MIT is plausibly to be a lock-in transition[39].

The development of "molecular orbitals"[40] or "orbital molecules"[41] in crystals has long been known and observed in a range of materials. Many materials, including the rutile and Magnéli phases[6,40,42], spinel $MgTi_2O_4$[43] and $CuIr_2S_4$[44] produce dimers. Three-orbitals molecules are uncommon, although regular triangular orbital molecules are seen in triangular lattice materials such as $LiVO_2$[45–47], $LiVS_2$[48], $AlV_2O_4$[49,50], and $BaV_{10}O_{15}$[51]. A two-center–two-electron (2c2e) bond is formed there, in which each of the three bonds is formed with two electrons, like a conventional chemical bond. In $LiVO_2$, for example, six electrons originating from three $V^{3+}$ ions participate in three bonds, each with two electrons[45–47].

Multi-centered bonds, such as three-centered–two-electron (3c2e) bonds, are present in electron-deficient compounds and clustered compounds like the Chevrel phase[38,52]. In $MTe_2$ (M = V, Nb, Ta)[53,54], $CrSe_2$[55], $Li_{0.33}VS_2$[56], structural transitions accompanied by the formation of linear trimers with 3c2e bonds are observed. For example, the linear trimer formation in $CrSe_2$ results in a 17% bond modulation in the Cr–Cr bond[55], with a displacement equivalent to RuP. The β-pyrochlore compound $CsW_2O_6$ forms regular triangular trimers at an MIT[57,58].

RuP is an electron-rich compound and is unique in that, to our knowledge, it is the first example of orbital trimers with 3c4e bonds; in comparison to 3c2e bonds, the extra two electrons go into non-bonding orbitals and does not contribute to the stabilization of the cluster, but neither does it destabilize it. Indeed, electron-rich molecules such as $I^{3-}$ and $HF^{2-}$ molecules are known to form 3c4e-bonded linear trimers[59]. In RuP, the linear trimer is formed by receiving electrons from the electron-rich phosphorus chain. Thus, the formation of 3c4e bond orbital clusters appears to be a common characteristic in stabilizing electron-rich compounds.

Finally, the melting of the linear trimer order is discussed: when the Ru site of RuP is substituted with Rh, the nonmagnetic insulator phase disappears at 10% substitution and the PG phase is suppressed more gradually approaching a QCP at 45%[21]. Interestingly, a dome-shaped superconducting phase appears around the QCP. The insulator and PG phases are suppressed because the trimer is destabilized by an increase in the number of electrons occupying the antibonding orbitals. Trimer fluctuations may thus play a role in superconductivity. Similar dome-shaped superconducting phases have been observed in $IrTe_2$[60,61] and $BaNi_2Ge_2$[62], when their dimers are suppressed by chemical substitutions. When the orbital cluster order is melted, the orbital fluctuation is expected to cause unique critical behavior and superconductivity.

**CONCLUSIONS**

In summary, we found that the polycrystalline sample of RuP exhibits an MIT while the single crystal sample of $RuP_{1-x}$ becomes a metal via electron doping owing to the phosphorous deficiency. The structure of the nonmagnetic insulator phase in the RuP polycrystalline sample was revealed by synchrotron XRD studies to be equivalent to that of RuAs with a 3×3×3 superlattice. The uniform Ru–Ru distance in the high-temperature phase is substantially modulated, resulting in the formation of a linear ruthenium trimer. The formation of a 3c4e bonded $d_{xy}$ orbital cluster induced by electron transfer from phosphorous is suggested to be responsible for the MIT of RuP.

**AUTHOR INFORMATION**

Corresponding Authors
*dhirai@nuap.nagoya-u.ac.jp
*hiroi@issp.u-tokyo.ac.jp

Present Addresses
†Department of Applied Physics, Nagoya University, Furo-cho, Chikusa-ku, Nagoya 464-8603, Japan.

Author Contributions

The manuscript was written through contributions of all authors. All authors have given approval to the final version of the manuscript.

Funding Sources

This work was partly supported by Japan Society for the Promotion of Science (JSPS) KAKENHI Grants of Numbers 20H01858, 22H04462, and 22H01178.

Notes
The authors declare no competing financial interest.

**ACKNOWLEDGMENT**

The authors are grateful to H. Kotegawa for insightful discussion and giving us structural parameters of RuAs at 170 K.

# Supplemental Materials

## a. Heat capacities of polycrystalline RuP and single crystal RuP$_{1-x}$

The heat capacities of a RuP polycrystalline sample and a RuP$_{1-x}$ single crystal are compared in Fig. S1. Both $C/T$ data are proportional to $T^2$ below ~8 K. Fits to the linear equation of $C/T = \gamma + \beta T^2$ yield $\gamma = 0.32(1)$ mJ K$^{-1}$ mol$^{-1}$ and $\beta = 0.033(1)$ mJ K$^{-4}$ mol$^{-1}$ for the polycrystalline sample and $\gamma = 1.23(1)$ mJ K$^{-1}$ mol$^{-1}$ and $\beta = 0.032(1)$ mJ K$^{-4}$ mol$^{-1}$ for the single crystal. The Debye temperatures derived from the $\beta$ values are 490 and 487 K for the polycrystalline and single crystal samples, respectively. In contrast to the nearly equal Debye temperatures, the Sommerfeld coefficients $\gamma$ are very different: the negligibly small $\gamma$ of the polycrystalline sample is consistent with the insulating resistivity, while the sizable $\gamma$ of the single crystal sample indicates a finite density of states at the Fermi energy, consistent with the metallic resistivity at low temperatures.

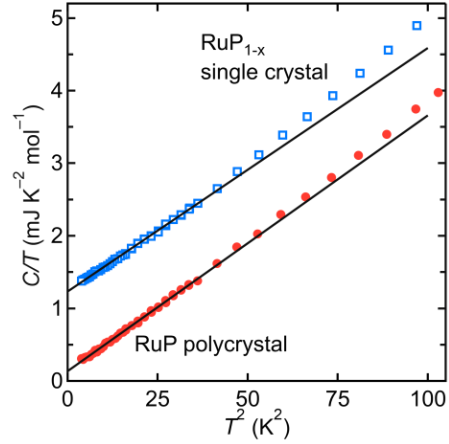

Fig. S1. Heat capacity divided by temperature ($C/T$) plotted against $T^2$ for a RuP polycrystalline sample (red filled circles) and a RuP$_{1-x}$ single crystal sample (blue open squares). The black lines are fits to the equation $C/T = \gamma + \beta T^2$.

## b. Hall coefficients of polycrystalline RuP and single crystal RuP$_{1-x}$

Hall coefficient $R_H(T)$ measurements were performed to examine the types and densities of carriers for RuP polycrystalline and RuP$_{1-x}$ single crystal samples (Fig. S2). The $R_H(T)$ of the former is always positive and increases exponentially with decreasing temperature below $T_{MI}$, which is consistent with the insulating behavior in resistivity. In contrast, the $R_H(T)$ of the latter sample changes its sign upon cooling: it is negative at 340 K, becomes positive at a $T$-independent value of $10^{-3}$ cm$^3$ C$^{-3}$ below $T_{MI}$, and then turns out to be negative below 25 K. The negative $R_H(T)$ in the metallic phase at high temperatures indicates that the P deficiency in RuP$_{1-x}$ increases electron carriers. In contrast to the polycrystalline sample, $R_H(T)$ of the single crystal remains small below $T_{MI}$, implying that the metallic state persists even below $T_{MI}$. The non-linear $B$-dependence of $\rho_{yx}(B)$ at low temperatures, as well as the sign change of $R_H(T)$ below 25 K, indicating that RuP$_{1-x}$ is semimetallic below $T_{MI}$.

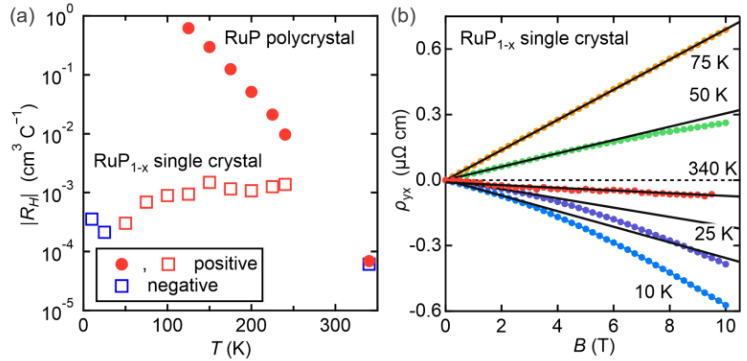

Fig. S2. (a) Temperature dependence of absolute value of Hall coefficients $|R_H(T)|$ for a polycrystalline RuP sample (filed circle) and a RuP$_{1-x}$ single crystal (open square). Positive and negative $R_H(T)$ are shown in red and blue, respectively. (b) Magnetic field dependence of Hall resistivity $\rho_{yx}(B)$ for a RuP$_{1-x}$ single crystal at different temperatures. Solid lines show a linear fit for $\rho_{yx}(B)$ in the low-field limit.

## c. Rietveld refinements

Powder X-ray diffraction (XRD) measurements were carried out at 400, 280, and 100 K using a synchrotron with the wavelength of $\lambda$ = 0.414 Å at beam line 11-BM at the Advanced Photon Source in the Argonne National Laboratory. A polycrystalline sample with the stoichiometric composition of RuP was examined. The crystal structure was refined by the Rietveld method using the RIETAN-FP program [1]. The XRD pattern at 400 K is well reproduced starting from the previously reported MnP-type structural model with the orthorhombic *Pnma* (Np. 62) space group [2,3], as shown in the Fig. S3. The refinement data are summarized in Table S1. The XRD pattern at 100 K was refined starting from the structural model reported for RuAs in the low temperature nonmagnetic insulating phase with the monoclinic $P2_1/c$ (No. 14) space group [4]. Good fits were obtained as shown in Fig. 4. Table S2 summarizes the crystallographic and Rietveld refinement data at 100 K. The crystal structure at 100 K is shown in Fig. S5. In the low temperature insulating phase, the Ru interatomic distance is varied to nine distinct lengths from the uniform Ru interatomic distance along the *b* axis in the high temperature metallic phase, as listed in Table S3. Six of them becomes shorter approximately 5%, while the other three becomes longer approximately 10%. As shown in Fig. S5, three Ru atoms are connected by two shorter bonds to create a linear trimer and are separated by one longer bond.

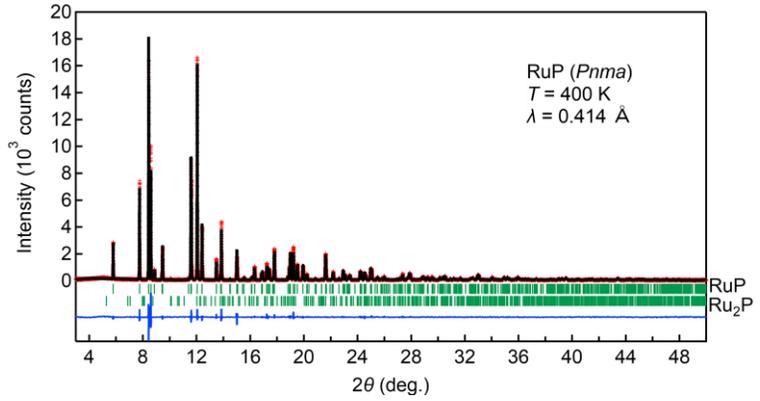

Fig. S3. Synchrotron powder X-ray diffraction pattern of polycrystalline RuP measured at 400 K shown with Rietveld refinements. Observed data, calculated line, and difference are shown by red crosses, black line, and lower blue line, respectively. Tick marks indicate the position of allowed reflections. Ru$_2$P is included as an impurity phase.

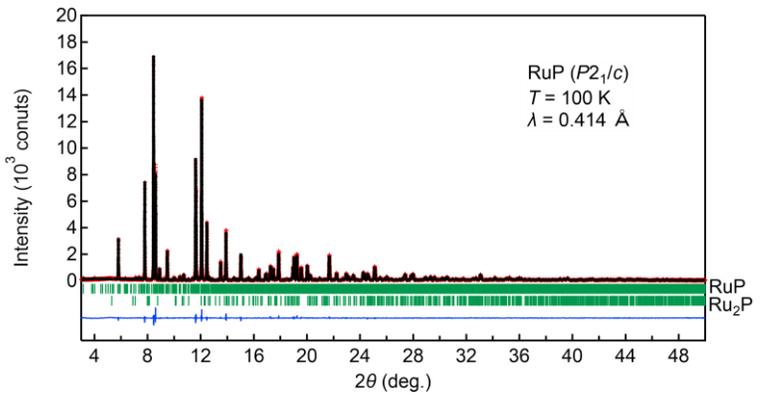

Fig. S4. Synchrotron powder X-ray diffraction pattern of polycrystalline RuP measured at 100 K shown with Rietveld refinements. Observed data, calculated line, and difference are shown by red crosses, black line, and lower blue line, respectively. Tick marks indicate the position of allowed reflections. Ru$_2$P is included as an impurity phase.

Table S1. Crystallographic and Rietveld refinement data for RuP in the metallic phase at 400 K. The equivalent isotropic atomic displacement parameter $B$ and the occupancy are also shown.

| Chemical formula: RuP | | | | | |
|---|---|---|---|---|---|
| Symmetry: *Pnma* (No. 62) | | | | | |
| $a$ = 5.52736(2) Å, $b$ = 3.17185(1) Å, $c$ = 6.12665(2) Å | | | | | |
| Site | Wyckoff | $x$ | $y$ | $z$ | Occupancy | $B$(Å$^2$) |
| Ru1 | 4$c$ | -0.00559(5) | 0.25 | 0.20501(3) | 1 | 0.636 |
| P1 | 4$c$ | 0.19054(13) | 0.25 | 0.56468(13) | 1 | 0.473 |
| Rietveld agreement factors | | | | | |
| $R_{wp}$ = 10.299 %, $R_p$ = 8.262 %, $R_e$ = 7.188 %, $S$ = 1.4328 | | | | | |

Table S2. Crystallographic and Rietveld refinement data for RuP in the insulating phase at 100 K. The equivalent isotropic atomic displacement parameter $B$ and the occupancy are also shown.

| Chemical formula: RuP | | | | | | |
|---|---|---|---|---|---|---|
| Symmetry: $P2_1/c$ (No. 14) | | | | | | |
| $a$ = 6.36892(2) Å, $b$ = 18.29734(5) Å, $c$ = 8.40037(3) Å, $\beta$ = 101.2032(2)° | | | | | | |
| Site | Wyckoff | $x$ | $y$ | $z$ | Occupancy | $B$(Å$^2$) |
| Ru1 | 4$e$ | -0.2281 | 0.92968 | 0.2655 | 1 | 0.128 |
| Ru2 | 4$e$ | 0.5634 | 0.76169 | 0.5656 | 1 | 0.128 |
| Ru3 | 4$e$ | 0.607 | 0.43625 | 0.6064 | 1 | 0.128 |
| Ru4 | 4$e$ | 0.1017 | 0.59364 | 0.595 | 1 | 0.128 |
| Ru5 | 4$e$ | -0.2716 | 0.89685 | 0.73 | 1 | 0.128 |
| Ru6 | 4$e$ | 0.082 | 0.9292 | 0.583 | 1 | 0.128 |
| Ru7 | 4$e$ | 0.2465 | 0.76553 | 0.2499 | 1 | 0.128 |
| Ru8 | 4$e$ | -0.0696 | 0.72771 | 0.4374 | 1 | 0.128 |
| Ru9 | 4$e$ | 0.4221 | 0.59671 | 0.9163 | 1 | 0.128 |
| P10 | 4$e$ | -0.3007 | 0.6385 | 0.5188 | 1 | 0.08 |
| P11 | 4$e$ | 0.64 | 0.692 | 0.8041 | 1 | 0.08 |
| P12 | 4$e$ | 0.05 | 0.8529 | 0.3562 | 1 | 0.08 |
| P13 | 4$e$ | -0.0621 | 0.8069 | 0.6475 | 1 | 0.08 |
| P14 | 4$e$ | 0.394 | 0.5219 | 0.6808 | 1 | 0.08 |
| P15 | 4$e$ | 0.9682 | 0.4741 | 0.6524 | 1 | 0.08 |
| P16 | 4$e$ | 0.365 | 0.8538 | 0.6776 | 1 | 0.08 |
| P17 | 4$e$ | 0.2833 | 0.6809 | 0.4697 | 1 | 0.08 |
| P18 | 4$e$ | -0.2871 | 0.9826 | 0.5178 | 1 | 0.08 |
| Rietveld agreement factors | | | | | | |
| $R_{wp}$ = 7.726 %, $R_p$ = 6.280 %, $R_e$ = 7.369 %, $S$ = 1.0484 | | | | | | |

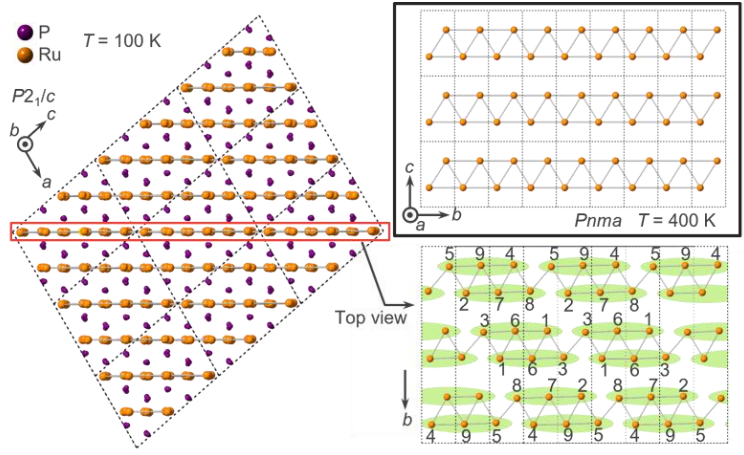

Fig. S5. Crystal structure of RuP in the insulating phase at 100 K. A set of unit cells viewed along the *b* axis are indicated by dotted lines. A section of the structure highlighted by a red square is viewed along the vertical direction and is shown in the lower right, which demonstrates the modulation of Ru interatomic distances in the chains. The distinct Ru sites are identified by numbers given in Table S2. Green ellipses represent linear trimers formed by shorter Ru bonds. In the upper right, the corresponding section of the crystal structure in the high temperature metallic phase is presented for comparison.

Table S3. The selected distances between Ru ions in the insulating phase at 100 K. The differences $\Delta d$ of the Ru inter atomic distances from 3.172 Å at 400 K are also listed.

| Sites | Distance (Å) | $\Delta d$ | Sites | Distance (Å) | $\Delta d$ |
|---|---|---|---|---|---|
| Ru1, Ru6 | 2.995(4) | −5.58% | Ru3, Ru1 | 3.559(3) | +12.20% |
| Ru6, Ru3 | 2.956(3) | −6.81% | Ru5, Ru4 | 3.500(4) | +10.34% |
| Ru4, Ru9 | 3.052(4) | −3.78% | Ru2, Ru8 | 3.524(4) | +11.10% |
| Ru9, Ru5 | 2.961(4) | −6.65% | | | |
| Ru8, Ru7 | 2.984(4) | −5.93% | | | |
| Ru7, Ru2 | 3.005(4) | −5.26% | | | |

### d. First-principles calculations

First-principles calculations were performed based on density functional theory (DFT), using the program package Quantum ESPRESSO [5], which employs plane waves and pseudopotentials to describe the Kohn–Sham orbitals and the crystalline potential, respectively. The plane-wave cutoff for a wave function was set to 55 Ry, which is suitable for the optimized norm-conserving pseudopotentials [6] provided in the SG15 pseudopotential library [7]. Calculations were performed with the Perdew-Burke-Ernzerhof functional [8] based on the generalized gradient correction. We perform the Brillouin-zone integral on 6×12×6 (6×2×4) *k*-point grids for the high- (low-) temperature phase by using the optimized tetrahedron method [9]. We computed the partial density of states by projecting each Kohn-Sham orbitals onto the maximally localized Wannier functions [10] calculated with the RESPACK [11] program package.